\documentclass[MSNbibl,nameyear,seceqn,dvips]{arxstspdf}
\usepackage{dcolumn}
\usepackage{flushend}
\usepackage{stfloats}
\usepackage{graphicx}

\volume{29}
\issue{2}
\pubyear{2014}
\firstpage{285}
\lastpage{301}
\doi{10.1214/13-STS455} %kopijuoti is PTS
\makeatletter
\def\textitt{}

\newcolumntype{d}[1]{D{.}{.}{#1}}
\renewcommand{\citep}[1]{\citeauthor{#1}, \citeyear{#1}}

\newtheorem{lem}{Lemma}

\newcommand{\eqref}[1]{(\ref{#1})}
\newcommand{\cov}{\operatorname{cov}}
\newcommand{\var}{\operatorname{var}}
\newcommand{\diag}{\operatorname{diag}}
\newcommand{\sd}{\operatorname{sd}}
\newcommand{\cv}{\operatorname{cv}}
\newcommand{\fdr}{\operatorname{fdr}}
\newcommand{\ufdr}{\operatorname{ufdr}}

\newcommand\caln{\mathcal{N}}

\newcommand\tref[1]{Theorem~\ref{#1}}

\newcommand\bmf{\mathbf{f}}
\newcommand\bmff{\mathbf{f}}
\newcommand\be{\mathbf{e}}
\newcommand\bg{\mathbf{g}}
\newcommand\bgg{\mathbf{g}}
\newcommand\bp{\mathbf{p}}
\newcommand\bq{\mathbf{q}}
\newcommand\bt{\mathbf{t}}
\newcommand\bU{\mathbf{U}}
\newcommand\bV{\mathbf{V}}
\newcommand\bX{\mathbf{X}}
\newcommand\by{\mathbf{y}}
\newcommand\bW{\mathbf{W}}

\newcommand\bzer{\mathbf{0}}
\newcommand\bone{\mathbf{1}}

\newcommand\bthe{\bolds\theta}

\newcommand\hate{\hat{E}}
\newcommand\hatf{\hat{f}}
\newcommand\hbf{\hat{\bmf}}
\newcommand\hbg{\hat{\bg}}
\newcommand\halp{\hat{\alpha}}
\newcommand\hdel{\hat{\Delta}}
\newcommand\hpi{\hat{\pi}}
\newcommand\hthe{\hat{\theta}}
\newcommand\pthe{(\theta)}

\newcommand\bare{\bar{E}}
\newcommand\barg{\bar{g}}

\newcommand\hfdr{\widehat{\fdr}}
\newcommand\hufdr{\widehat{\ufdr}}

\newcommand\iid{\stackrel{\mathrm{i.i.d.}}{\sim}}
\newcommand\ind{\stackrel{\mathrm{ind}}{\sim}}
\newcommand\pdot{(\cdot)}

\makeatother

\begin{document}
\begin{frontmatter}

\title{Two Modeling Strategies for Empirical Bayes Estimation}%
% kai straipsnis turi susijusiu diskusiju ir rejoinder'iu
%rejoinder at \relateddoi{r}{10.1214/00-STSXXXX}.}
\runtitle{EB Modeling, Computation and Accuracy}

\begin{aug}
\author[a]{\fnms{Bradley} \snm{Efron}\corref{}\ead[label=e1]{brad@stat.stanford.edu}}
\runauthor{B. Efron}

\affiliation{Stanford University}

\address[a]{Bradley Efron is Professor of Statistics and Biostatistics,
Department of Statistics, Stanford University, Stanford, California 94305-4065, USA \printead{e1}.}

\end{aug}

% ABSTRACT
%
\begin{abstract}
Empirical Bayes methods use the data from parallel experiments, for
instance, observations $X_k\sim\mathcal{N}(\Theta_k,1)$ for
$k=1,2,\ldots
,N$, to estimate the conditional distributions $\Theta_k|X_k$. There
are two main estimation strategies: modeling on the $\theta$ space,
called ``$g$-modeling'' here, and modeling on the $x$ space, called
``$f$-modeling.'' The two approaches are described and compared. A
series of computational formulas are developed to assess their
frequentist accuracy. Several examples, both contrived and genuine,
show the strengths and limitations of the two strategies.
\end{abstract}

% KEYWORDS

\end{frontmatter}

%s1 #&#
\section{Introduction}\label{sec1}

Empirical Bayes methods, though of increasing use, still suffer from an
uncertain theoretical basis, enjoying neither the safe haven of Bayes
theorem nor the steady support of frequentist optimality. Their
rationale is often reduced to inserting more or less obvious estimates
into familiar Bayesian formulas. This conceals the essential empirical
Bayes task: learning an appropriate prior distribution from ongoing
statistical experience, rather than knowing it by assumption. Efficient
learning requires both Bayesian and frequentist modeling strategies. My
plan here is to discuss such strategies in a mathematically simplified
framework that, hopefully, renders them more transparent. The
development proceeds with some methodological discussion supplemented
by numerical examples.

A wide range of empirical Bayes applications have the following
structure: repeated sampling from an unknown prior distribution $g\pthe
$ yields unseen realizations
%
%e1.1 #&#
\begin{equation}
\Theta_1,\Theta_2,\ldots,\Theta_N.
\label{11}
\end{equation}
Each $\Theta_k$ in turn provides an observation $X_k\sim f_{\Theta
_k}\pdot$ from a known probability family $f_\theta(x)$,
%
%e1.2 #&#
\begin{equation}
X_1,X_2,\ldots,X_N. \label{12}
\end{equation}
On the basis of the observed sample \eqref{12}, the statistician wishes
to approximate certain Bayesian inferences that would be directly
available if $g\pthe$ were known. This is the empirical Bayes framework
developed and named by \citet{robbins}. Both $\Theta$ and $X$ are
usually one-dimensional variates, as they will be in our examples,
though that is of more applied than theoretical necessity.

A central feature of empirical Bayes estimation is that the data
arrives on the $x$ scale but inferences are calculated on the $\theta$
scale. Two main strategies have developed: modeling on the $\theta$
scale, called \textit{$g$-modeling} here, and modeling on the $x$
scale, called \textit{$f$-modeling}. $G$-modeling has predominated in
the theoretical empirical Bayes literature, as in \citet{laird}, \citet
{morris}, \citet{zhang}, and \citet{jiang}. Applications, on the other
hand, from \citet{robbins} onward, have more often relied on
$f$-modeling, recently as in \citeauthor{2010} (\citeyear{2010,2011}) and \citet{brown}.

We begin Section~\ref{sec2} with a discretized statement of Bayes theorem that
simplifies the nonparametric\vadjust{\goodbreak} $f$-modeling development of Section~\ref{sec3}.
Parameterized $f$-modeling, necessary for efficient empirical Bayes
estimation, is discussed in Section~\ref{sec4}. Section~\ref{sec5} introduces an
exponential family class of $g$-modeling procedures. Classic empirical
Bayes applications, an $f$-modeling stronghold (including Robbins'
Poisson formula, the James--Stein estimator and false discovery rate
methods), are the subject of Section~\ref{sec6}. The paper concludes with a
brief discussion in Section~\ref{sec7}.

Several numerical examples, both contrived and genuine, are carried
through in Sections~\ref{sec2} through \ref{sec7}. The comparison is
never one-sided: as
one moves away from the classic applications, $g$-modeling comes into
its own. Trying to go backward, from observations on the $x$-space to
the unknown prior $g\pthe$, has an ill-posed computational flavor.
Empirical Bayes calculations are inherently fraught with difficulties,
making both of the modeling strategies useful. An excellent review of
empirical Bayes methodology appears in Chapter~3 of \citet{newcarlin}.

There is an extensive literature, much of it focusing on rates of
convergence, concerning the ``deconvolution problem,'' that is,
estimating the distribution $g\pthe$ from the observed $X$ values. A
good recent reference is \citet{butucea}. Empirical Bayes inference
amounts to estimating certain nonlinear functionals of $g\pdot$,
whereas linear functionals play a central role for the deconvolution
problem, as in \citet{cavalier}, but the two literatures are related.
The development in this paper employs discrete models that avoid rates
of convergence difficulties.

Empirical Bayes analyses often produce impressive-looking estimates of
posterior $\theta$ distributions. The main results in what follows are
a series of computational formulas---Theorems \ref{th1} through \ref{th4}---giving
the accuracy of both $f$-model and $g$-model estimates. Accuracy can be
poor, as some of the examples show, and in any case accuracy
assessments are an important part of the analysis.

%s2 #&#
\section{A Discrete Model of Bayesian Inference}\label{sec2}

In order to simplify the $f$-modeling computations, we will assume a
model in which both the parameter vector $\theta$ and the observed data
set $x$ are confined to finite discrete sets:
%
%e2.1 #&#
\begin{eqnarray}\label{21}
\theta\in\bthe&=&(\theta_1,\theta_2,\ldots,
\theta_j,\ldots,\theta_m)\quad\mbox{and}
\nonumber
\\[-8pt]
\\[-8pt]
\nonumber
x\in\mathbf{x}&=&(x_1,x_2,
\ldots,x_i,\ldots,x_n)
\end{eqnarray}
with $m<n$. The prior distribution $\bg$ puts probability $g_j$ on
$\theta_j$,
%
%e2.2 #&#
\begin{equation}
\bg=(g_1,g_2,\ldots,g_j,
\ldots,g_m)'. \label{22}
\end{equation}
This induces a marginal distribution $\bmf$ on $\mathbf{x}$,
%
%e2.3 #&#
\begin{equation}
\bmf=(f_1,f_2,\ldots,f_i,
\ldots,f_n)', \label{23}
\end{equation}
with $f_i=\Pr\{x=x_i\}$. Letting $\{p_{ij}\}$ represent the sampling
probabilities
%
%e2.4 #&#
\begin{equation}
p_{ij}=\Pr\{x_i|\theta_j\}, \label{24}
\end{equation}
the $n\times m$ matrix
%
%e2.5 #&#
\begin{equation}
P=(p_{ij}) \label{25}
\end{equation}
produces $\bmf$ from $\bg$ according to
%
%e2.6 #&#
\begin{equation}
\bmf=P\bg. \label{26}
\end{equation}

%f1 #&#
\begin{figure*}

\includegraphics{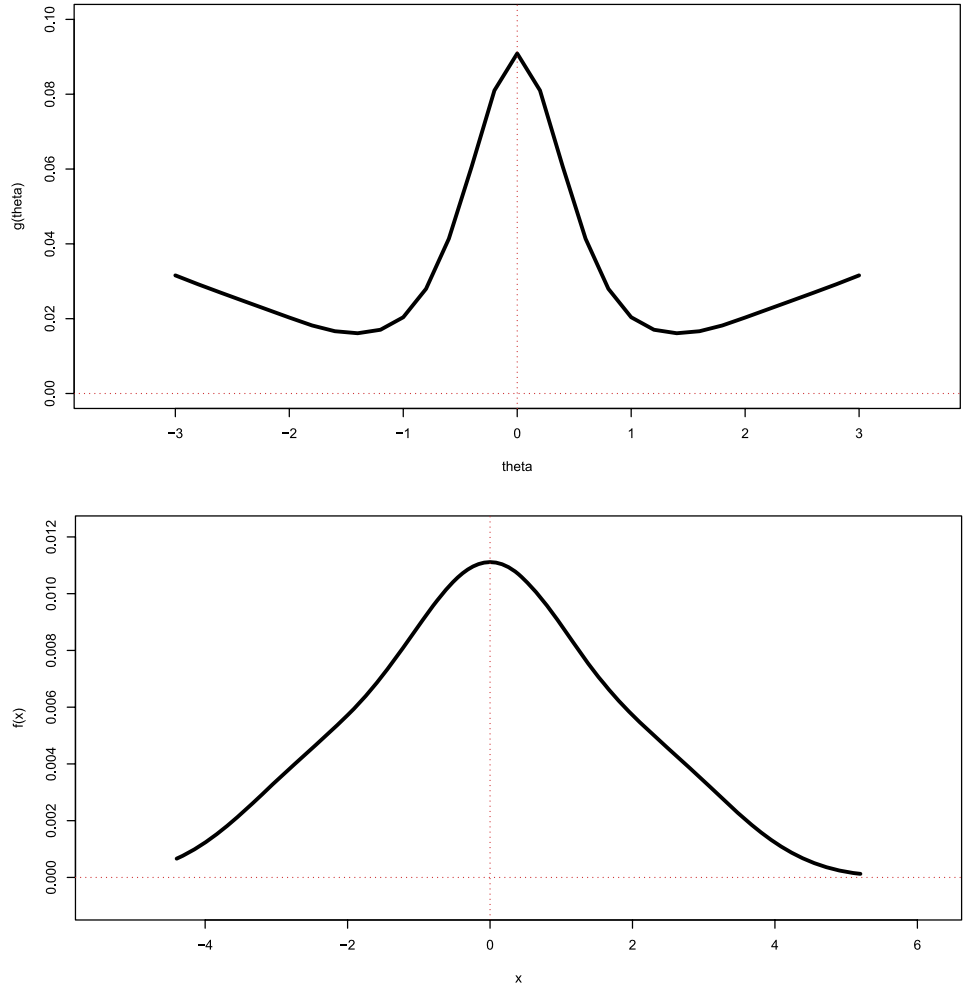}

\caption{\emph{Top:} Discrete model: prior $g\pthe, \theta
=\operatorname
{seq}(-3,3,0.2)$; $g$ is equal mixture of $\caln(0,0.5^2)$ and density
$\propto|\theta|$. \emph{Bottom:} Corresponding $f(x)$: assuming
$\caln(\theta,1)$ sampling, $x=\operatorname{seq}(-4.4,5.2,0.05)$.
Note the
different scales.}
\label{fig1}
\end{figure*}

In the example of Figure~\ref{fig1}, we have
%
%e2.7 #&#
\begin{equation}
\bthe=(-3,-2.8,\ldots,3)\quad (m=31), \label{27}
\end{equation}
with $g\pthe$ an equal mixture of a discretized $\caln(0,0.5^2)$
density and a density proportional to $|\theta|$. The sampling
probabilities $p_{ij}$ are obtained from the normal translation model
$\varphi(x_i-\theta_j)$, $\varphi$ the standard normal density
function, and with
%
%e2.8 #&#
\begin{equation}
\mathbf{x}=(-4.4,-4.35,\ldots,5.2)\quad (n=193). \label{28}
\end{equation}
Then $\bmf=P\bg$ produces the triangular-shaped mar\-ginal density $f(x)$
seen in the bottom panel. Looking ahead, we will want to use samples
from the bottom distribution to estimate functions of the top.

In the discrete model \eqref{21}--\eqref{26}, Bayes rule takes the form
%
%e2.9 #&#
\begin{equation}
\Pr\{\theta_j|x_i\}=p_{ij}g_j/f_i.
\label{29}
\end{equation}
Letting $\bp_i$ represent the $i$th row of matrix $P$, the $m$-vector
of posterior probabilities of $\theta$ given $x=x_i$ is given by
%
%e2.10 #&#
\begin{equation}
\diag(\bp_i)\bg/\bp_i\bg, \label{210}
\end{equation}
where $\diag(\mathbf{v})$ indicates a diagonal matix with diagonal elements
taken from the vector $\mathbf{v}$.

Now suppose $t\pthe$ is a parameter of interest, expressed in our
discrete setting by the vector of values
%
%e2.11 #&#
\begin{equation}
\bt=(t_1,t_2,\ldots,t_j,
\ldots,t_m)'. \label{211}
\end{equation}
The posterior expectation of $t\pthe$ given $x=x_i$ is then
%
%e2.12 #&#
\begin{eqnarray}
\label{212} %
E \{t\pthe|x_i \}&=&\sum
_{j=1}^mt_jp_{ij}g_j\Big/f_i
\nonumber
\\[-8pt]
\\[-8pt]
\nonumber
&=&\bt'\diag(\bp_i)\bg/\bp_i\bg.
\end{eqnarray}

The main role of the discrete model \eqref{21}--\eqref{26} is to
simplify the presentation of $f$-modeling begun in Section~\ref{sec3}.
Basically, it allows the use of familiar matrix calculations rather
than functional equations. $G$-modeling, Section~\ref{sec5}, will be presented
in both discrete and continuous forms. The prostate data example of
Section~\ref
{sec6} shows our discrete model nicely handling continuous data.

%s3 #&#
\section{Bayes Rule in Terms of \lowercase{$\mathbf{f}$}}\label{sec3}

Formula \eqref{212} expresses $E\{t\pthe|x_i\}$ in terms of the prior
distribution $\bg$. This is fine for pure Bayesian applications but in
empirical Bayes work, information arrives on the $x$ scale and we may
need to express Bayes rule in terms of $\bmf$. We begin by inverting
\eqref{26}, $\bmf=P\bg$.

For now assume that the $n\times m$ matrix $P$ \eqref{24}--\eqref{25}
is of full rank $m$. Then the $m\times n$ matrix
%
%e3.1 #&#
\begin{equation}
A=\bigl(P'P\bigr)^{-1}P' \label{31}
\end{equation}
carries out the inversion,
%
%e3.2 #&#
\begin{equation}
\bg=A\bmf. \label{32}
\end{equation}
Section~\ref{sec4} discusses the case where rank($P$) is less than~$m$. Other
definitions of $A$ are possible; see the discussion in Section~\ref{sec7}.

With $\bp_i$ denoting the $i$th row of $P$ as before, let
%
%e3.3 #&#
\begin{equation}\quad
\mathbf{u}'=(\cdots t_jp_{ij}\cdots)=
\bt'\diag(\bp_i), \quad\mathbf{v}'=
\bp_i \label{33}
\end{equation}
and
%
%e3.4 #&#
\begin{equation}
\bU'=\mathbf{u}'A, \quad\bV'=\mathbf{v}'A,
\label{34}
\end{equation}
$\bU$ and $\bV$ being $n$-vectors. (Here we are suppressing the
subscript in $\mathbf{u}=\mathbf{u}_i$, etc.) Using \eqref{32}, the Bayes posterior
expectation $E\{t|x_i\}$ \eqref{212} becomes
%
%e3.5 #&#
\begin{equation}
E\{t|x_i\}=\frac{\mathbf{u}'\bg}{\mathbf{v}'\bg}=\frac{\bU'\bmf}{\bV
'\bmf}, \label{35}
\end{equation}
the latter being \textit{Bayes rule in terms of $\bmf$}. Notice that
$\bU$ and $\bV$ do not depend on $\bg$ or $\bmf$. The denominator
$\bV
'\bmf$ equals $f(x_i)$ in \eqref{35}, but not in the regularized
versions of Section~\ref{sec4}.

In a typical empirical Bayes situation, as in Section~6.1 of \citet
{2010}, we might observe independent observations $X_1,X_2,\ldots,X_N$
from the marginal density $f(x)$,
%
%e3.6 #&#
\begin{equation}
X_k\iid f\pdot,\quad k=1,2,\ldots,N, \label{36}
\end{equation}
and wish to estimate $E=E\{t|x_i\}$. For the discrete model \eqref{21},
the vector of counts $\by=(y_1,y_2,\ldots,y_n)'$,
%
%e3.7 #&#
\begin{equation}
y_i=\#\{X_k=x_i\}, \label{37}
\end{equation}
is a nonparametric sufficient statistic; $\by$ follows a multinomial
distribution on $n$ categories, $N$ draws, probability vector $\bmf$,
%
%e3.8 #&#
\begin{equation}
\by\sim\operatorname{Mult}_n(N,\bmf), \label{38}
\end{equation}
having mean vector and covariance matrix
%
%e3.9 #&#
\begin{equation}
\by\sim\bigl(N\bmf,ND(\bmf) \bigr), \quad D(\bmf)\equiv\diag(\bmf
)-\bmf
\bmf'. \label{39}
\end{equation}

The unbiased estimate of $\bmf$,
%
%e3.10 #&#
\begin{equation}
\hbf=\by/N, \label{310}
\end{equation}
gives a nonparametric estimate $\hate$ of $E\{t|x_i\}$ by substitution
into \eqref{35},
%
%e3.11 #&#
\begin{equation}
\hate=\bU'\hbf/\bV'\hbf. \label{311}
\end{equation}

Using $\hbf\sim(\bmf,D(\bmf)/N)$, a standard differential argument
yields the approximate ``delta method'' frequentist standard error of
$\hate$. Define
%
%e3.12 #&#
\begin{equation}
U_f=\sum_{i=1}^nf_iU_i,\quad
V_f=\sum_{i=1}^nf_iV_i
\label{312}
\end{equation}
and
%
%e3.13 #&#
\begin{equation}
\bW=\frac{\bU}{U_f}-\frac{\bV}{V_f}. \label{313}
\end{equation}
(Notice that $\sum f_iW_i=0$.)
%
%th1 #&#
\begin{thm}\label{th1}
The delta-method approximate standard deviation of $\hate=\bU'\hbf
/\bV
'\hbf$ is
%
%e3.14 #&#
\begin{equation}
\sd(\hate)=\frac{1}{\sqrt{N}}|E|\cdot\sigma_f(W), \label{314}
\end{equation}
where $E=\bU'\bmf/\bV'\bmf$ and
%
%e3.15 #&#
\begin{equation}
\sigma_f^2(W)=\sum_{i=1}^nf_iW_i^2.
\label{315}
\end{equation}
The approximate coefficient of variation $\sd(\hate)/|E|$ of $\hate$ is
%
%e3.16 #&#
\begin{equation}
\cv(\hate)=\sigma_f(W) /\sqrt{N}. \label{316}
\end{equation}
\end{thm}
\begin{pf}
From \eqref{35} we compute the joint moments of $\bU'\hbf$ and $\bV
'\hbf$,
%
%e3.17 #&#
\begin{eqnarray}\label{317}
\qquad&&\pmatrix{\bU'\hbf
\cr
\bV'\hbf}
\nonumber
\\[-8pt]
\\[-8pt]
\nonumber
&&\quad\sim\biggl(
\pmatrix{U_f
\cr
V_f},\frac{1}{N} %
\pmatrix{\sigma_f^2(U)&\sigma_f(U,V)
\cr
\sigma_f(U,V)&\sigma_f^2(V) } %
\biggr),
\end{eqnarray}
with $\sigma_f^2(U)=\sum f_i(U_i-U_f)^2, \sigma_f(U,V)= \sum
f_i(U_i-U_f)(V_i-V_f)$, and $\sigma_f^2(V)=\sum f_i(V_i-V_f)^2$. Then
%
%e3.18 #&#
\begin{eqnarray}
\label{318} %
\hate=\frac{\bU'\hbf}{\bV'\hbf}&=&E\cdot\frac{1+\hdel
_U}{1+\hdel_V}\quad
\nonumber\\
&\doteq& E\cdot(1+\hdel_U-\hdel_V ),\\
\eqntext{\displaystyle\biggl[\hdel_U=\frac{\bU'\hbf-U_f}{U_f}, \hdel_V=
\frac{\bV'\hbf
-V_f}{V_f} \biggr]}
\end{eqnarray}
so $\sd(\hate^2)\doteq E^2\var(\hdel_U-\hdel_V)$, which, again using~\eqref{39}, gives \tref{th1}.
\end{pf}

The trouble here, as will be shown, is that $\sd(\hate)$ or $\cv
(\hate
)$ may easily become unmanageably large. Empirical Bayes methods
require sampling on the $x$ scale, which can be grossly inefficient for
estimating functions of $\theta$.

Hypothetically, the $X_k$'s in \eqref{36} are the observable halves of
pairs $(\Theta,X)$,
%
%e3.19 #&#
\begin{equation}\qquad
(\Theta_k,X_k)\ind g\pthe f_\theta(x),\quad  k=1,2,
\ldots,N. \label{319}
\end{equation}
If the $\Theta_k$'s \textit{had} been observed, we could estimate
$\bg$
directly as $\bar{\bg}=(\barg_1,\barg_2,\ldots,\barg_m)'$,
%
%e3.20 #&#
\begin{equation}
\barg_j=\#\{\Theta_k=\theta_j\}/N,
\label{320}
\end{equation}
leading to the \textit{direct Bayes estimate}
%
%e3.21 #&#
\begin{equation}
\bare=\mathbf{u}'\bar{\bg}/\mathbf{v}'\bar{\bg}.
\label{321}
\end{equation}

$\bare$ would usually be less variable than $\hate$ \eqref{311} (and
would automatically enforce possible constraints on $E$ such as
monotonicity in $x_k$). A version of \tref{th1} applies here. Now we define
%
%e3.22 #&#
\begin{eqnarray}
u_g&=&\sum_{j=1}^mg_ju_j,\quad
v_g=\sum_{j=1}^mg_jv_j\quad
\mbox{and}
\nonumber
\\[-8pt]
\\[-8pt]
\nonumber
\mathbf{w}&=&\mathbf{u} /u_g-\mathbf{v}/v_g. \label{322}
\end{eqnarray}

%th2 #&#
\begin{thm}\label{th2}
For direct Bayes estimation \eqref{321}, the delta-method approximate
standard deviation of $\bare$ is
%
%e3.23 #&#
\begin{equation}
\sd(\bare)=\frac{1}{\sqrt{N}}|E|\cdot\sigma_g(w), \label{323}
\end{equation}
where
%
%e3.24 #&#
\begin{equation}
\sigma_g^2(w)=\sum_{j=1}^mg_jw_j^2;
\label{324}
\end{equation}
$\bare$ has approximate coefficient of variation
%
%e3.25 #&#
\begin{equation}
\cv(\bare)=\sigma_g(w) /\sqrt{N}. \label{325}
\end{equation}
\end{thm}
The proof of \tref{th2} is the same as that for \tref{th1}.

%t1 #&#
\begin{table}
\tabcolsep=0pt
\caption{Standard deviation and coefficient of variation of $E\{t\pthe
|x=2.5\}$ (for $N=1$); for the three parameters \protect\eqref{326},
with $g$
and $f$ as in Figure~\protect\ref{fig1}; sdf from \protect\tref{th1}
\protect\eqref{314}; sdd for
direct Bayes estimation, \protect\tref{th2} \protect\eqref{323};
sdx from the
regularized $f$-modeling of Section~\protect\ref{sec4}, \protect\tref{thm3}
\protect\eqref{48}}%
\label{tab1}
\begin{tabular*}{\columnwidth}{@{\extracolsep{\fill
}}lcd{2.2}d{2.2}d{2.2}ccc@{}}
\hline
&&\multicolumn{3}{c}{$\bolds{N^{1/2}}$ \textbf{sd}}&
\multicolumn{3}{c}{$\bolds{N^{1/2}}$ \textbf{cv}}\\[-6pt]
&&\multicolumn{3}{c}{\hrulefill}&
\multicolumn{3}{c@{}}{\hrulefill}\\
$\bolds{t\pthe}$ &\multicolumn{1}{c}{$\bolds{E\{t|x=2.5\}
}$}&\multicolumn{1}{c}{\textbf{sdf}}&
\multicolumn{1}{c}{\textbf{sdd}} &
\multicolumn{1}{c}{\textbf{sdx}} & \multicolumn{1}{c}{\textbf{cvf}} &
\multicolumn{1}{c}{\textbf{cvd}}& \multicolumn{1}{c@{}}{\textbf
{cvx}}\\
\hline
Parameter (1)& 2.00 & 8.74& 3.38& 2.83& 4.4& 1.7& 1.4\\
Parameter (2)& 4.76 & 43.4& 13.7& 10.4& 9.1& 2.9& 2.2\\
Parameter (3)& 0.03 & 43.9& 0.53& 1.24& 1371& 16& 39\\
\hline
\end{tabular*}
\end{table}
Table~\ref{tab1} concerns the estimation of $E\{t\pthe|x=2.5\}$ for the
situation shown in Figure~\ref{fig1}.\vadjust{\goodbreak} Three different parameters $t\pthe$ are
considered:
%
%e3.26 #&#
\begin{eqnarray}\label{326}
(1)\quad &t\pthe=\theta,
\nonumber\\
(2)\quad &t\pthe=\theta^2,
\\
(3)\quad &t\pthe= %
\cases{1,&$\mbox{if }\theta\leq0,$\vspace*{2pt}
\cr
0,&$\mbox{if }\theta>0$.}\nonumber
\end{eqnarray}
In the third case, $E\{t\pthe|x\}=\Pr\{\theta\leq0|x\}$. \textitt{Cvf}
is $\sqrt{N}\cv(\hate)$ \eqref{316} so cvf$/\sqrt{N}$ is the
approximate coefficient of variation of $\hate$, the nonparametric
empirical Bayes estimate of $E\{t\pthe|x=2.5\}$. \textitt{Cvd} is the
corresponding quantity \eqref{325}, available only if we could directly
observe the $\Theta_k$ values in \eqref{319}, while \textitt{cvx} is a
regularized version of $\hate$ described in the next section.

Suppose we wish to bound $\cv(\hate)$ below some prespecified value
$c_0$, perhaps $c_0=0.1$. Then according to \eqref{316}, we need $N$
to equal
%
%e3.27 #&#
\begin{equation}
N=(\cv_1/c_0)^2, \label{327}
\end{equation}
where $\cv_1$ is the numerator $\sigma_f(W)$ of \eqref{316}, for
example, cvf in Table~\ref{tab1}. For the three parameters \eqref{326} and
for $c_0=0.1$, we would require $N=1936$, 8281 and 187 million, respectively.

%f2 #&#
\begin{figure*}

\includegraphics{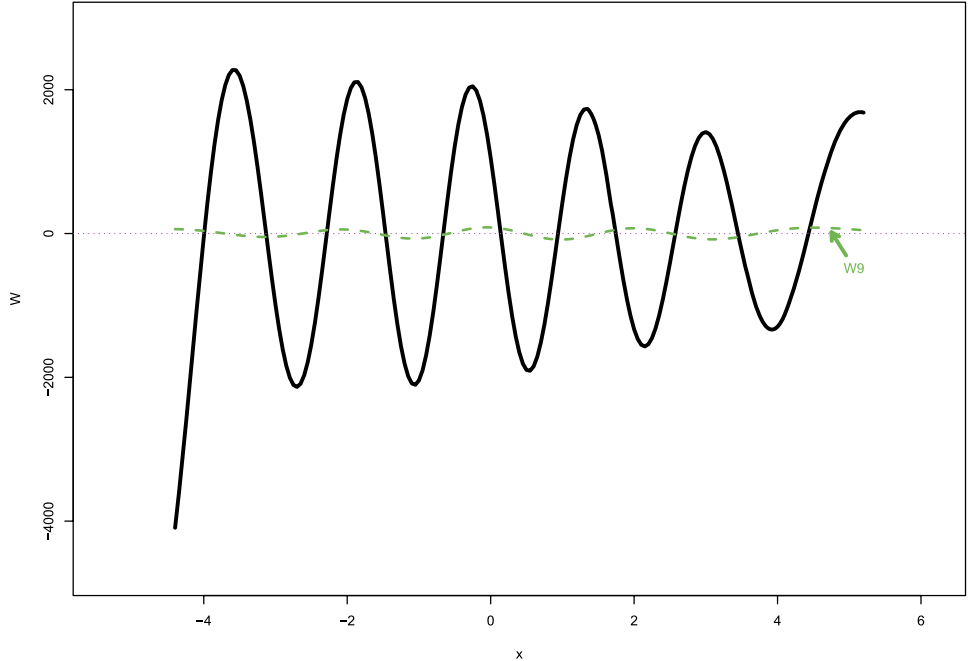}

\caption{$\bW$ vector \protect\eqref{313} for $f$-Bayes estimation
of $\Pr\{
\theta\leq0|x=2.5\}$ for the model of Figure~\protect\ref{fig1} (actually
$\bW_{12}$
as in Section~\protect\ref{sec4}; dashed curve is $\bW_9$).}
\label{fig2}
\end{figure*}

The vector $\bW$ for parameter (3) is seen to take on enormous values
in Figure~\ref{fig2}, resulting in $\sigma_f(W)=1370.7$ for \eqref{316}. The
trouble stems from the abrupt discontinuity of $t_3$ at $\theta=0$,
which destabilizes $\bU$ in \eqref{313}. Definition \eqref{34} implies
$\bU'P=\mathbf{u}'$. This says that $\bU'$ must linearly compose $\mathbf
{u}'$ from
the rows of $P$. But in our example the rows of $P$ are smooth
functions of the form $\varphi(x_i-\theta_j)$, forcing the violent
cycling of $U$ seen in Figure~\ref{fig2}. Section~\ref{sec4} discusses a
regularization method that greatly improves the accuracy of using
``Bayes rule in terms of $\bmf$.''

Table~\ref{tab1} shows that if we \textit{could} sample on the $\theta$
scale, as in \eqref{320}, we would require ``only'' 25,600 $\Theta_k$
observations to achieve coefficient of variation 0.1 for estimating
$\Pr
\{\theta\leq0|x=2.5\}$; direct sampling is almost always more efficient
than $f$ sampling, but that is not the way empirical Bayes situations
present themselves. The efficiency difference is a factor of 86 for
parameter (3), but less than a factor of 3 for parameter (1), $t\pthe
=\theta$. The latter is a particularly favorable case for empirical
Bayes estimation, as discussed in Section~\ref{sec6}.

The assumption of independent sampling, \eqref{36} and \eqref{319}, is
a crucial element of all our results. Independence assumptions (often
tacitly made) dominate the empirical Bayes literature, as in \citet
{muralidharan}, \citet{zhang}, \citet{morris}, and \citet{1975morris}.
Nonindependence effectively reduces the effective sample size $N$; see
Chapter~8 of \citet{2010}. This point is brought up again in Section~\ref{sec6}.

%s4 #&#
\section{Regularized \lowercase{$f$}-Modeling}\label{sec4}

Fully nonparametric estimation of $E=E\{t\pthe|x\}$ is sometimes
feasible, but, as seen in Table~\ref{tab1} of Section~\ref{sec3}, it can become
unacceptably noisy. Some form of regularization is usually necessary. A
promising approach is to estimate $\bmf$ parametrically according to a
smooth low-dimensional model.

Suppose then that we have such a model, yielding $\hbf$ as an estimate
of $\bmf$ \eqref{23}, with mean vector and covariance matrix
%
%e4.1 #&#
\begin{equation}
\hbf\sim\bigl(\bmf,\Delta(\bmf)/N \bigr). \label{41}
\end{equation}
In the nonparametric case \eqref{39} $\Delta(\bmf)=D(\bmf)$, but we
expect that we can reduce $\Delta(\bmf)$ parametrically. In any case,
the delta-method approximate coefficient of variation for $\hate=\bU
'\hbf/\bV'\hbf$ \eqref{311} is given in terms of $\bW$~\eqref{313}:
%
%e4.2 #&#
\begin{equation}
\cv(\hate)= \bigl\{\bW'\Delta(\bmf)\bW/N \bigr\}^{1/2}.
\label{42}
\end{equation}
This agrees with \eqref{316} in the nonparametric situation~\eqref{39}
where $\Delta(\bmf)=\diag(\bmf)-\bmf\bmf'$. The verification of
\eqref
{42} is almost identical to that for \tref{th1}.

Poisson regression models are convenient for the smooth parametric
estimation of $\bmf$. Beginning with an $n\times p$ structure matrix
$\bX$, having rows $\mathbf{x}_i$ for $i=1,2,\ldots,n$, we assume that the
components of the count vector $\by$ \eqref{37} are independent Poisson
observations,
%
%e4.3 #&#
\begin{eqnarray}\label{43}
y_i\ind\operatorname{Poi}(\mu_i),\quad \mu_i=e^{\mathbf{x}_i\alpha}
\nonumber
\\[-8pt]
\\[-8pt]
\eqntext{\mbox{for }i=1,2,\ldots,n,}
\end{eqnarray}
where $\alpha$ is an unknown vector of dimension $p$. Matrix~$\bX$ is
assumed to have as its first column a vector of 1's.

Let $\mu_+=\sum_1^n\mu_i$ and $N=\sum_1^ny_i$, and define
%
%e4.4 #&#
\begin{equation}
f_i=\mu_i/\mu_+\quad \mbox{for }i=1,2,\ldots,n. \label{44}
\end{equation}
Then a well-known Poisson/multinomial relationship says that the
conditional distribution of $\by$ given $N$ is
%
%e4.5 #&#
\begin{equation}
\by|N\sim\operatorname{Mult}_n(N,\bmf) \label{45}
\end{equation}
as in \eqref{38}. Moreover, under mild regularity conditions, the
estimate $\hbf=\by/N$ has asymptotic mean vector and covariance matrix
(as $\mu_+\to\infty$)
%
%e4.6 #&#
\begin{equation}
\hbf\,\dot\sim\,\bigl(\bmf,\Delta(\bmf)/N \bigr), \label{46}
\end{equation}
where
%
%e4.7 #&#
\begin{eqnarray}\label{47}
\Delta(\bmf)=\diag(\bmf)\bX G_f^{-1}\bX'
\diag(\bmf)
\nonumber
\\[-8pt]
\\[-8pt]
\eqntext{\bigl[G_f=\bX'\diag(\bmf)\bX\bigr];}
\end{eqnarray}
Equations~\eqref{46}--\eqref{47} are derived from standard generalized linear
model calculations. Combining \eqref{42} and~\eqref{46} gives a Poisson
regression version of \tref{th1}.
%
%th3 #&#
\begin{thm}\label{th3}
The delta-method coefficient of variation for $\hate=\bU'\hbf/\bV
'\hbf$
under Poisson model \eqref{43} is
%
%e4.8 #&#
\begin{equation}\quad
\cv(\hate)= \bigl\{\bigl(\bW'\bX\bigr)_f\bigl(
\bX'\bX\bigr)_f^{-1}\bigl(\bW'\bX
\bigr)_f'/N \bigr\}^{1/2}, \label{48}
\end{equation}
where
%
%e4.9 #&#
\begin{eqnarray}\label{49}
\bigl(\bW'\bX\bigr)_f&=&\bW'\diag(\bmf)\bX\quad
\mbox{and}
\nonumber
\\[-8pt]
\\[-8pt]
\nonumber
\bigl(\bX'\bX\bigr)_f&=&\bX'\diag(
\bmf)\bX,
\end{eqnarray}
with $\bW$ as in \eqref{313}.
\label{thm3}
\end{thm}

The bracketed term in \eqref{48}, times $N$, is recognized as the
length$^2$ of the projection of $\bW$ into the $p$-dimensional space
spanned by the columns of $\bX$, carried out using inner product
$\langle a,b\rangle_f=\sum f_ia_ib_i$. In the nonparametric case, $\bX$
equals the identity $I$, and \eqref{48} reduces to \eqref{316}. As in
\eqref{314}, $\sd(\hate)$ is approximated by $|E|\cv(\hate)$.
[\textit
{Note}: \tref{thm3} remains valid as stated if a multinomial model for
$\hbf$ replaces the Poisson calculations in \eqref{47}.]

\textitt{Cvx} in Table~\ref{tab1} was calculated as in \eqref{48}, with $N=1$.
The structure matrix $\bX$ for the example in Figure~\ref{fig1} was obtained
from the R natural spline function $ns(x,df=5)$; including a column of
1's made $\bX193\times6$. The improvements over \textitt{cvf}, the
nonparametric coefficients of variation, were by factors of 3, 5 and
100 for the three parameters \eqref{326}.

The regularization in \tref{thm3} takes place with respect to $\bmf$
and $\hbf$. Good performance also requires regularization of the
inversion process $\hbg=A\hbf$ \eqref{32}. Going back to the beginning
of Section~\ref{sec3}, let
%
%e4.10 #&#
\begin{equation}
P=LDR' \label{410}
\end{equation}
represent the singular value decomposition of the $n\times m$ matrix
$P$, with $L$ the $n\times m$ orthonormal matrix of left singular
vectors, $R$ the $m\times m$ orthonormal matrix of right singular
vectors, and $D$ the $m\times m$ diagonal matrix of singular values,
%
%e4.11 #&#
\begin{equation}
d_1\geq d_2\geq\cdots\geq d_m.
\label{411}
\end{equation}
Then it is easy to show that the $m\times n$ matrix
%
%e4.12 #&#
\begin{equation}
A=RD^{-1}L' \label{412}
\end{equation}
is the \textit{pseudo-inverse} of $P$, which is why we could go from
$\bmf=P\bg$ to $\bg=A\bmf$ at \eqref{32}. [Other pseudo-inverses exist;
see \eqref{71}.]

Definition \eqref{412} depends on $P$ being of full rank~$m$,
equivalently having $d_m>0$ in \eqref{411}. Whether or not this is
true, very small values of $d_j$ will destabilize $A$.\vadjust{\goodbreak} The familiar
cure is to truncate representation~\eqref{412}, lopping off the end
terms of the singular value decomposition. If we wish to stop after the
first $r$ terms, we define $R_r$ to be the first $r$ columns of $R$,
$L_r$ the first $r$ columns of $L$, $D_r$ the $r\times r$ diagonal
matrix $\diag(d_1,d_2,\ldots,d_r)$, and
%
%e4.13 #&#
\begin{equation}
A_r=R_rD_r^{-1}L_r'.
\label{413}
\end{equation}
In fact, $r=12$ was used in Figure~\ref{fig2} and Table~\ref{tab1}, chosen to make
%
%e4.14 #&#
\begin{equation}
\sum_{r+1}^md_j^2
\bigg/\sum_1^md_j^2<10^{-10}.
\label{414}
\end{equation}

As in \eqref{31}--\eqref{313}, let
%
%e4.15 #&#
\begin{equation}
\bU_r'=\mathbf{u}'A_r,\quad
\bV_r'=\mathbf{v}'A_r
\label{415}
\end{equation}
[$\mathbf{u}$ and $\mathbf{v}$ stay the same as in (\ref{33})],
%
%e4.16 #&#
\begin{equation}
E_r=\frac{\bU_r'\bmf}{\bV_r'\bmf},\quad \hate_r=\frac{\bU_r'\hbf
}{\bV
_r'\hbf}
\label{416}
\end{equation}
and
%
%e4.17 #&#
\begin{equation}
\bW_r=\frac{\bU_r}{\sum f_iU_{ri}}-\frac{\bV_r}{\sum f_iV_{ri}}.
\label{417}
\end{equation}
\tref{thm3} then remains valid, with $\bW_r$ replacing $\bW$.
\textit
{Note}: Another regularization method, which will not be pursued here,
is the use of ridge regression rather than truncation in the inversion
process \eqref{32}, as in \citet{hall}.

%t2 #&#
\begin{table}[b]
\tabcolsep=0pt
\caption{Coefficient of variation and standard deviation ($N=1$), for
$E\{t|x=2.5\}$ as in \protect\ref{tab1}; now using Poisson regression
in \protect\tref
{thm3}, with $\bX$ based on a natural spline with 5 degrees of freedom.
Increasing choice of $r$, \protect\eqref{413}--\protect\eqref{417},
decreases bias but
increases variability of $\hate$ for parameter (3); $g$ error from
\protect\eqref{420}}\label{tab2}
\begin{tabular*}{\columnwidth}{@{\extracolsep{\fill}}lcccccd{6.1}d{5.1}@{}}
\hline
&&\multicolumn{3}{c}{\textbf{Parameter (1)}}&
\multicolumn{3}{c}{\textbf{Parameter (3)}}\\[-6pt]
&&\multicolumn{3}{c}{\hrulefill}&\multicolumn{3}{c}{\hrulefill}\\
$\bolds{r}$&$\bolds{g}$ \textbf{error}& $\bolds{E_r}$ & \textbf
{cvx} &
\textbf{sdx}&$\bolds{E_r}$&\multicolumn{1}{c}{\textbf{cvx}} &
\multicolumn{1}{c@{}}{\textbf{sdx}}\\
\hline
\phantom{0}3& 0.464& 1.75& 1.00& 1.75& 0.021& 3.6& 0.1\\
\phantom{0}6& 0.254& 2.00& 1.34& 2.68& 0.027& 4.6& 0.1\\
\phantom{0}9& 0.110& 2.00& 1.36& 2.73& 0.031& 8.2& 0.3\\
12& 0.067& 2.00& 1.41& 2.83& 0.032& 38.6& 1.2\\
15& 0.024& 2.00& 1.39& 2.78& 0.033& 494.0& 16.1\\
18& 0.012& 2.00& 1.39& 2.78& 0.033& 23\mbox{,}820.8& 783.8\\
21& 0.006& 2.00& 1.40& 2.80& 0.033&960\mbox{,}036.4&31\mbox{,}688.8\\
\hline
\end{tabular*}
\end{table}

%f3 #&#
\begin{figure*}

\includegraphics{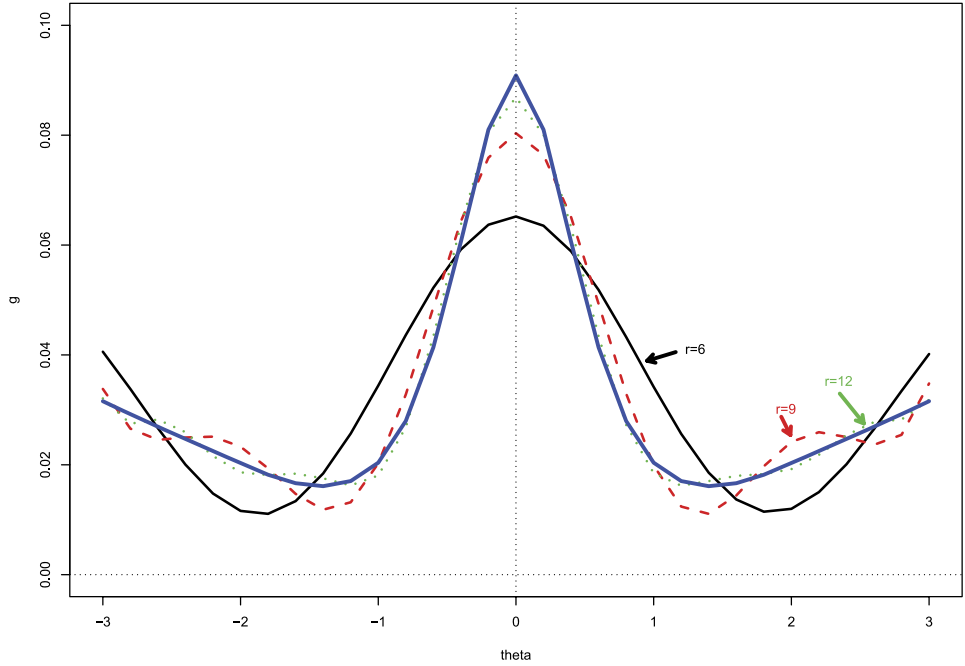}

\caption{Approximation $g_r$ \protect\eqref{418} with $r=6,9,12$ for
$g$ in Figure~\protect\ref
{fig1}; heavy blue curve is $g$.}
\label{fig3}
\end{figure*}

Reducing $r$ reduces $\bW_r$, hence reducing \eqref{49} and the
approximate coefficient of variation of $\hate_r$. The reduction can be
dramatic. $W_9$ almost disappears compared to $W_{12}$ in Figure~\ref{fig2}.
Table~\ref{tab2} compares various choices of $r$ for parameters (1) and (3)
\eqref{326}. The choice turns out to be unimportant for parameter (1)
and crucial for parameter (3).

Why not always choose a small value of $r$? The trouble lies in
possible bias for the estimation of $E=E\{t|x\}$. Rather than the
crucial inverse mapping $\bg=A\bmf$ \eqref{32}, we get an approximation
%
%e4.18 #&#
\begin{eqnarray}
\label{418} %
\bg_r&=&A_r
\bmf=A_rP\bg
\nonumber
\\[-8pt]
\\[-8pt]
\nonumber
&=&R_rD_r^{-1}L_r'LDR'
\bg=R_rR_r'\bg%
\end{eqnarray}
[the last step following from $LDR'=L_rD_rR_r'+L_{(r)}D_{(r)}R_{(r)}'$,
with $L_{(r)}$ indicating the last $m-r$ columns of $L$, etc.; Equation~\eqref
{418} says that $\bg_r$ is the projection of $\bg$ into the linear
space spanned by the first $r$ columns of $R$]. Then, looking at \eqref
{415}--\eqref{416},
%
%e4.19 #&#
\begin{equation}
E_r=\frac{\bU_r'\bmf}{\bV_r'\bmf}=\frac{\mathbf{u}'\bg_r}{\mathbf
{v}'\bg_r}, \label{419}
\end{equation}
possibly making $\hate_r$ badly biased for estimating $E=\mathbf{u}'\bg
/\mathbf{v}'\bg$.

The $E_r$ columns of Table~\ref{tab2} show that bias is a problem only for
quite small values of $r$. However, the example of Figure~\ref{fig1} is
``easy'' in the sense that the true prior $\bg$ is smooth, which allows
$\bg_r$ to rapidly approach $\bg$ as $r$ increases, as pictured in
Figure~\ref
{fig3}. The $g_{\mathrm{error}}$ column of Table~\ref{tab2} shows this
numerically in terms of the absolute error
%
%e4.20 #&#
\begin{equation}
g_{\mathrm{error}}=\sum_{i=1}^m|g_{ri}-g_i|.
\label{420}
\end{equation}

\begin{figure*}

\includegraphics{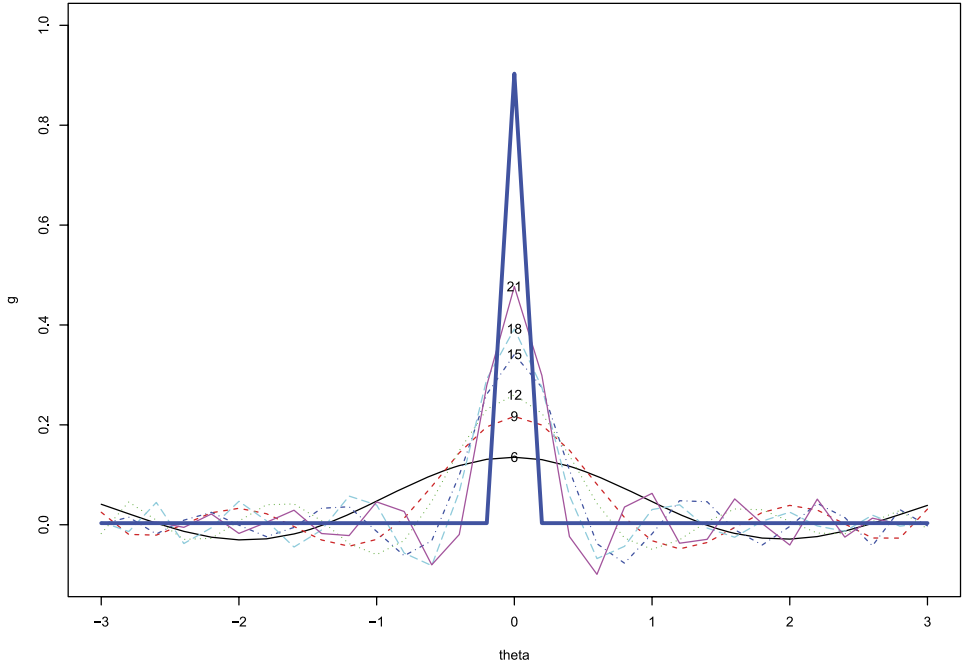}

\caption{True $g=0.90\cdot\delta(0)+0.10$ uniform (heavy curve);
approximation $g_r$ \protect\eqref{418} for $r=6,9,12,15,18,21$, as labeled.}
\label{fig4}
\end{figure*}

A more difficult case is illustrated in Figure~\ref{fig4}. Here $\bg$ is a
mixture: 90\% of a delta function at $\theta=0$ and 10\% of a uniform
distribution over the 31 points $\theta_j$ in $\bthe=(-3,-2.8,\ldots
,3)$; $P$ and $\mathbf{x}$ are as before. Now $g_{\mathrm{error}}$
exceeds 1.75
even for $r=21$; $\bg_r$ puts too small a weight on $\theta=0$, while
bouncing around erratically for $\theta\neq0$, often going negative.
%
%f4 #&#

We expect, correctly, that empirical Bayes estimation of $E\{t\pthe|x\}
$ will usually be difficult for the situation of Figure~\ref{fig4}. This is
worrisome since its $\bg$ is a reasonable model for familiar false
discovery rate analyses, but see Section~\ref{sec6}. Section~\ref{sec5} discusses a
different regularization approach that ameliorates, without curing, the
difficulties seen here.

%s5 #&#
\section{Modeling the Prior Distribution \lowercase{$\mathbf{g}$}}\label{sec5}

The regularization methods of Section~\ref{sec4} involved modeling $\bmf$, the
marginal distribution \eqref{23} on the $x$-space, for example, by
Poisson regression in Table~\ref{tab2}. Here we discuss an alternative
strategy: modeling $\bg$, the prior distribution \eqref{22} on the
$\theta$-space. This has both advantages and disadvantages, as will be
discussed.

We begin with an $m\times q$ model matrix $Q$, $j$th row $Q_j$, which determines $\bg$
according to
%
%e5.1 #&#
\begin{equation}\qquad
\bg(\alpha)=e^{Q\alpha-\bone_m\phi(\alpha)} \quad\Biggl[\phi(\alpha
)=\log\sum
_1^me^{Q_j\alpha} \Biggr]. \label{51}
\end{equation}
[For $\mathbf{v}=(v_1,v_2,\ldots,v_m), e^{\mathbf{v}}$ denotes a vector with
components $e^{v_j}$; $\bone_m$ is a vector of $m$ 1's, indicating in
\eqref{51} that $\phi(\alpha)$ is subtracted from each component of
$Q\alpha$.] Here $\alpha$ is the unknown $q$-dimensional natural
parameter of exponential family \eqref{51}, which determines the prior
distribution $\bg=\bg(\alpha)$. In an empirical Bayes framework,
$\bg$
gives $\bmf=P\bg$ \eqref{26}, and the statistician then observes a
multinomial sample $\by$ of size $N$ from $\bmf$ as in \eqref{38},
%
%e5.2 #&#
\begin{equation}
\by\sim\mathrm{Mult}_n \bigl(N,P\bg(\alpha) \bigr), \label{52}
\end{equation}
from which inferences about $\bg$ are to be drawn.

Model \eqref{51}--\eqref{52} is not an exponential family in $\by$,
a~theoretical disadvantage compared to the Poisson modeling of \tref
{thm3}. [It is a \textit{curved exponential family}, \citet{1975}.]
We can still pursue an asymptotic analysis of its frequentist accuracy. Let
%
%e5.3 #&#
\begin{equation}
D(\bg)\equiv\diag(\bg)-\bg\bg', \label{53}
\end{equation}
the covariance matrix of a single random draw $\Theta$ from
distribution $\bg$, and define
%
%e5.4 #&#
\begin{equation}
Q_\alpha=D \bigl(\bg(\alpha) \bigr)Q. \label{54}
\end{equation}

%le1 #&#
\begin{lem}
The Fisher information matrix for estimating $\alpha$ in model \eqref
{51}--\eqref{52} is
%
%e5.5 #&#
\begin{equation}
\mathcal{I}=NQ_\alpha'P'\diag\bigl(1/\bmf(
\alpha) \bigr)PQ_\alpha, \label{55}
\end{equation}
where $P$ is the sampling density matrix \eqref{25}, and $\bmf(\alpha
)=P\bg(\alpha)$.
\label{lem1}
\end{lem}
\begin{pf}
Differentiating $\log\bg$ in \eqref{51} gives the $m\times q$
derivative matrix $d\log g_i/d\alpha_k$,
%
%e5.6 #&#
\begin{equation}
\frac{d\log\bg}{d\alpha}= \bigl[I-\bone_m\bg(\alpha)'
\bigr]Q, \label{56}
\end{equation}
so
%
%e5.7 #&#
\begin{eqnarray}
\label{57} %
\frac{d\bg}{d\alpha}&=&\diag\bigl(\bg(\alpha) \bigr)
\frac{d\log
\bg
}{d\alpha}
\nonumber
\\[-8pt]
\\[-8pt]
\nonumber
&=&D \bigl(\bg(\alpha) \bigr)Q=Q_\alpha.
\end{eqnarray}
This yields $d\bmf/d\alpha=PQ_\alpha$ and
%
%e5.8 #&#
\begin{equation}
\frac{d\log\bmf}{d\alpha}=\diag\biggl(\frac{1}{\bmf(\alpha
)} \biggr)PQ_\alpha.
\label{58}
\end{equation}

%f5 #&#
\begin{figure*}

\includegraphics{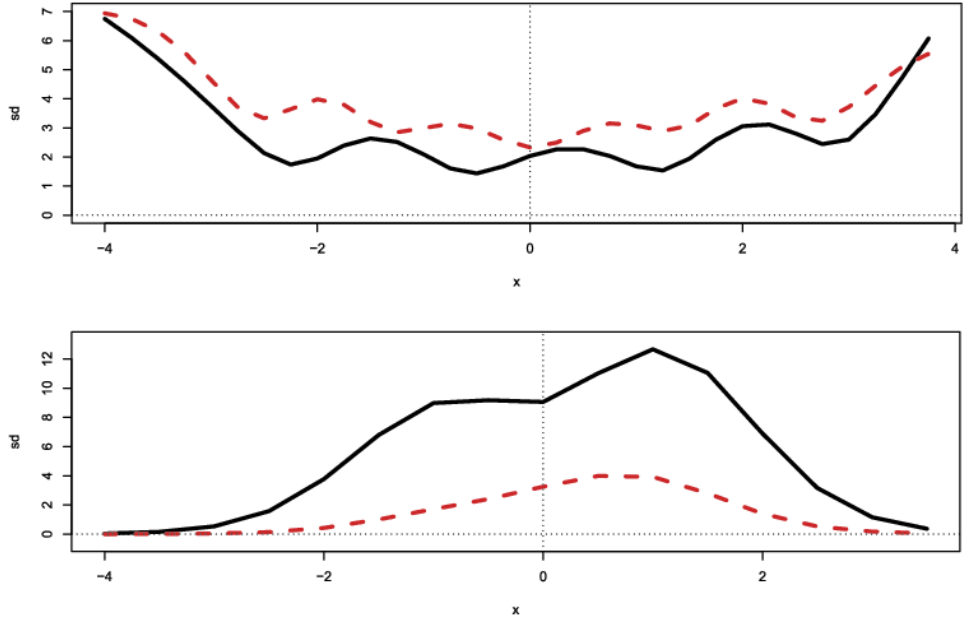}

\caption{\emph{Top:} Standard deviation of $E\{t|x\}$ as a
function of $x$, for parameter (1) $t\pthe=\theta$ (with $N=1$);
$f$-modeling (solid), $g$-modeling (dashed). \emph{Bottom:} Now
for parameter (3), $t\pthe=1$ or 0 as $\theta\leq0$ or $>0$; using
natural spline models, $df=6$, for both calculations.}
\label{fig5}
\end{figure*}

The log likelihood from multinomial sample \eqref{52} is
%
%e5.9 #&#
\begin{equation}
l_\alpha(\by)=\by'\log\bmf(\alpha)+\mathrm{ constant},
\label{59}
\end{equation}
giving score vector
%
%e5.10 #&#
\begin{equation}
\frac{dl_\alpha(\by)}{d\alpha}=\by'\frac{d\log\bmf}{d\alpha}.
\label{510}
\end{equation}
Since $\by$ has covariance matrix $N(\diag\bmf-\bmf\bmf')$ \eqref{39},
$\mathcal{I}$, the covariance matrix of the score vector, equals
%
%e5.11 #&#
\begin{eqnarray}
\label{511} %
\mathcal{I}&=&NQ_\alpha'P'
\diag(1/\bmf) \bigl(\diag\bmf-\bmf\bmf'\bigr)\nonumber\\
&&{}\cdot \diag(1/\bmf
)PQ_\alpha
\\
&=&NQ_\alpha'P' \bigl(\diag(1/\bmf)-
\bone_n\bone_n' \bigr)PQ_\alpha.\nonumber
\end{eqnarray}
Finally,
%
%e5.12 #&#
\begin{equation}
\bone_n'PQ_\alpha=\bone_m'D
\bigl(g(\alpha)\bigr)Q=\bzer'Q=0 \label{512}
\end{equation}
(using the fact that the columns of $P$ sum to 1), and \eqref{511}
yields the lemma.
\end{pf}

Standard sampling theory says that the maximum likelihood estimate
(MLE) $\halp$ has approximate covariance matrix $\mathcal{I}^{-1}$ and
that $\hbg=\bg(\halp)$ has approximate covariance, from \eqref{57},
%
%e5.13 #&#
\begin{equation}
\cov(\hbg)=Q_\alpha\mathcal{I}^{-1}Q_\alpha'.
\label{513}
\end{equation}

%le2 #&#
\begin{lem}\label{lem2}
The approximate covariance matrix for the maximum likelihood estimate
$\bg(\halp)$ of $\bg$ in model \eqref{51}--\eqref{52} is
%
%e5.14 #&#
\begin{eqnarray}\label{514}\qquad
&&\cov(\hbg)
\nonumber
\\[-8pt]
\\[-8pt]
\nonumber
&&\quad=\frac{1}{N}Q_\alpha\bigl[Q_\alpha'P'
\diag\bigl(1/\bmf(\alpha) \bigr)PQ_\alpha\bigr]^{-1}Q_\alpha'.
\end{eqnarray}
\end{lem}

If we are interested in a real-valued parameter $\tau=T(\bg)$, the
approximate standard deviation of its MLE $\hat\tau=T(g(\halp))$ is
%
%e5.15 #&#
\begin{equation}
\sd(\hat\tau)= \bigl[\dot{T}'\cov(\hbg)\dot{T}
\bigr]^{1/2}, \label{515}
\end{equation}
where $\dot{T}$ is the gradient vector $dT/d\bg$, evaluated at $\hbg$.
When $T(\bg)$ is the conditional expectation of a parameter $t\pthe$
\eqref{35},
%
%e5.16 #&#
\begin{equation}
T(\bg)=E \{t\pthe|x=x_i \}=\mathbf{u}'\bg/
\mathbf{v}'\bg, \label{516}
\end{equation}
we compute
%
%e5.17 #&#
\begin{equation}
\dot{T}(\bg)=\mathbf{w}=(\mathbf{u}/u_g)-(\mathbf{v}/v_g)
\label{517}
\end{equation}
\eqref{322}, and get the following.
%
%th4 #&#
\begin{thm}\label{th4}
Under models \eqref{51}--\eqref{52}, the MLE $\hate$ of $E\{t\pthe
|x=x_i\}$ has approximate standard deviation
%
%e5.18 #&#
\begin{equation}
\sd(\hate)=|E| \bigl[\mathbf{w}'\cov(\hbg)\mathbf{w} \bigr]^{1/2},
\label{518}
\end{equation}
with $\mathbf{w}$ as in \eqref{517} and $\cov(\hbg)$ from \eqref{514}.
\label{thm4}
\end{thm}

We can now compare $\sd(\hate)$ from $\bgg$-modeling \eqref{518}, with
the corresponding $\bmff$-modeling results of \tref{thm3}. Figure~\ref{fig5}
does this with parameters (1) and (3) \eqref{326} for the example of
Figure~\ref{fig1}. \tref{thm3}, modified as at \eqref{417} with $r=12$,
represents $\bmff$-modeling, now with $X$ based on $ns(\mathbf{x},6)$, a
natural spline with six degrees of freedom. Similarly for $\bgg
$-modeling, $Q=ns(\bthe,6)$ in \eqref{51}; $\alpha$ was chosen to make
$\bg(\alpha)$ very close to the upper curve in Figure~\ref{fig1}. (Doing so
required six rather than five degrees of freedom.)

The upper panel of Figure~\ref{fig5} shows $\bmff$-modeling yielding somewhat
smaller standard deviations for parameter (1), $t\pthe=\theta$. This is
an especially favorable case for $\bmff$-modeling, as discussed in Section~\ref
{sec6}. However, for parameter (3), $E=\Pr\{t\leq0|x\}$, $\bgg$-modeling
is far superior. \textit{Note:} in exponential families, curved or not,
it can be argued that the effective degrees of freedom of a model
equals its number of free parameters; see Remark D of \citet{2004}. The
models used in Figure~\ref{fig5} each have six parameters, so in this sense
the comparison is fair.

Parametric $g$-space modeling, as in \eqref{51}, has several advantages
over the $f$-space modeling of Section~\ref{sec4}:

\textit{Constraints}. $\hbg=\exp(Q\halp-\bone_m\phi(\halp))$ has
all coordinates positive, unlike the estimates seen in Figure~\ref{fig4}.
Other constraints such as monotonicity or convexity that may be imposed
on $\hbf=P\hbg$ by the structure of $P$ are automatically enforced, as
discussed in Chapter~3 of \citet{newcarlin}.

\textit{Accuracy}. With some important exceptions, discussed in
Section~\ref
{sec6}, $g$-modeling often yields smaller values of $\sd(\hate)$, as
typified in the bottom panel of Figure~\ref{fig5}. This is particularly true
for discontinuous parameters $t\pthe$, such as parameter (3) in Table~\ref{tab1}.

\textit{Simplicity}. The bias/variance trade-offs involved with the
choice of $r$ in Section~\ref{sec4} are avoided and, in fact, there is no need
for ``Bayes rule in terms of $\bmf$.''

%t3 #&#
\begin{table*}
\caption{Estimating $E=\Pr\{\theta=0|x\}$ in the situation of
Figure~\protect\ref
{fig4}; using $g$-modeling \protect\eqref{51} with $Q$ equal $ns(x,5)$
augmented with a column putting a delta function at $\theta=0$. Sd is
$\sd(\hate)$ \protect\eqref{525}, cv is the coefficient of
variation $\sd/E$.
(For sample size $N$, divide entries by $N^{1/2}$.)}\label{tab3}
\begin{tabular*}{\textwidth}{@{\extracolsep{\fill
}}ld{2.2}d{2.2}d{2.2}d{2.2}d{2.2}d{2.2}d{2.2}d{2.2}d{2.2}@{}}
\hline
$\bolds{x}$& \multicolumn{1}{c}{$\bolds{-4}$}&
\multicolumn{1}{c}{$\bolds{-3}$}&
\multicolumn{1}{c}{$\bolds{-2}$} &
\multicolumn{1}{c}{$\bolds{-1}$}&
\multicolumn{1}{c}{\textbf{0}} &
\multicolumn{1}{c}{\textbf{1}}&
\multicolumn{1}{c}{\textbf{2}}& \multicolumn{1}{c}{\textbf{3}}&
\multicolumn{1}{c@{}}{\textbf{4}}\\%
\hline
$E$& 0.04& 0.32& 0.78& 0.94& 0.96& 0.94& 0.78& 0.32& 0.04\\[3pt]
$N^{1/2}\cdot$ sd& 0.95& 3.28& 9.77& 10.64& 9.70& 10.48& 9.92& 3.36&
0.75\\
$N^{1/2}\cdot$ cv& 24.23& 10.39& 12.53& 11.38& 10.09& 11.20& 12.72&
10.65& 19.21\\
\hline
\end{tabular*}
\end{table*}

\textit{Continuous formulation}. It is straightforward to translate
$g$-modeling from the discrete framework \eqref{21}--\eqref{24} into
more familiar continuous language. Exponential family model \eqref{51}
now becomes
%
%e5.19 #&#
\begin{eqnarray}\label{519}
g_\alpha\pthe=e^{\bq\pthe\alpha-\phi(\alpha)}
\nonumber
\\[-8pt]
\\[-8pt]
\eqntext{\displaystyle \biggl[\phi
(\alpha)=\log\int
e^{\bq\pthe\alpha} \,d\theta\biggr], }
\end{eqnarray}
where $\bq\pthe$ is a smoothly defined $1\times q$ vector function of
$\theta$. Letting $f_\theta(x)$ denote the sampling density of $x$
given $\theta$, define
%
%e5.20 #&#
\begin{eqnarray}\label{520}
h(x)=\int f_\theta(x)g\pthe(\bq\pthe-\bar{\bq} ) \,d\theta
\nonumber
\\[-8pt]
\\[-8pt]
\eqntext{\displaystyle\biggl
[\bar{
\bq}=\int g\pthe\bq\pthe \,d\theta\biggr].}
\end{eqnarray}
Then the $q\times q$ information matrix $\mathcal{I}$ \eqref{55} is
%
%e5.21 #&#
\begin{eqnarray}\label{521}
\mathcal{I}=N\int\biggl[\frac{h(x)'h(x)}{f(x)^2} \biggr]f(x) \,dx
\nonumber
\\[-8pt]
\\[-8pt]
\eqntext{\displaystyle\biggl[f(x)=\int g
\pthe f_\theta(x) \,dx \biggr].}
\end{eqnarray}
A posterior expectation $E=E\{t\pthe|x\}$ has MLE
%
%e5.22 #&#
\begin{equation}\qquad\hspace*{2pt}
\hate=\int t\pthe f_\theta(x)g_{\halp}\pthe \,d\theta\Big/\int
f_\theta(x)g_{\halp}\pthe \,d\theta. \label{522}\hspace*{-6pt}
\end{equation}
An influence function argument shows that $E$ has gradient
%
%e5.23 #&#
\begin{equation}
\frac{dE}{d\alpha}=E\int z\pthe g_\alpha\pthe(\bq\pthe-\bar{\bq
} ) \,d
\theta, \label{523}
\end{equation}
with
%
%e5.24 #&#
\begin{eqnarray}\label{524}
z\pthe&=&\frac{t\pthe f_\theta(x)g_\alpha\pthe}{\int t(\varphi
)f_\varphi
(x)g_\alpha(\varphi) \,d\varphi}
\nonumber
\\[-8pt]
\\[-8pt]
\nonumber
&&{}-\frac{f_\theta(x)g_\alpha\pthe
}{\int
f_\varphi(x)g_\alpha(\varphi) \,d\varphi}.
\end{eqnarray}
Then the approximate standard deviation of $\hate$ is
%
%e5.25 #&#
\begin{equation}
\sd(\hate)= \biggl(\frac{dE}{d\alpha}\mathcal{I}^{-1}
\frac
{dE}{d\alpha
}' \biggr)^{1/2}, \label{525}
\end{equation}
combining \eqref{521}--\eqref{524}. [Of course, the integrals required
in \eqref{525} would usually be done numerically, implicitly returning
us to discrete calculations!]

\textit{Modeling the prior}. Modeling on the $g$-scale is
convenient for situations where the statistician has qualitative
knowledge concerning the shape of the prior~$\bg$. As a familiar
example, large-scale testing problems often have a big atom of prior
probability at $\theta=0$, corresponding to the null cases. We can
accommodate this by including in model matrix $Q$ \eqref{51} a column
$\be_0=(0,0,\ldots,0,1,0,\ldots,0)'$, with the 1 at $\theta=0$.

Such an analysis was carried out for the situation in Figure~\ref{fig4}, where
the true $\bg$ equaled $0.9\be_0+0.1\cdot\operatorname{uniform}$. $Q$ was taken to be
the natural spline basis $ns(\bthe,5)$ augmented by column $\be_0$, a
$31\times6$ matrix. Table~\ref{tab3} shows the results for $\mathbf{t}=\be_0$,
that is, for
%
%e5.26 #&#
\begin{equation}
E=E\{t|x\}=\Pr\{\theta=0|x\}. \label{526}
\end{equation}
The table gives $E$ and $\sd(\hate)$ \eqref{518} for $x=-4,-3,\break  \ldots
,4\ (N=1)$, as well as the coefficient of variation $\sd(\hate)/E$.

The results are not particularly encouraging: we would need sample
sizes $N$ on the order of 10,000 to expect reasonably accurate
estimates $\hate$ \eqref{327}. On the other hand, $f$-modeling as in
Section~\ref{sec4} is hopeless here. Section~\ref{sec6} has more to say about false
discovery rate estimates \eqref{526}.

%f6 #&#
\begin{figure*}[b]

\includegraphics{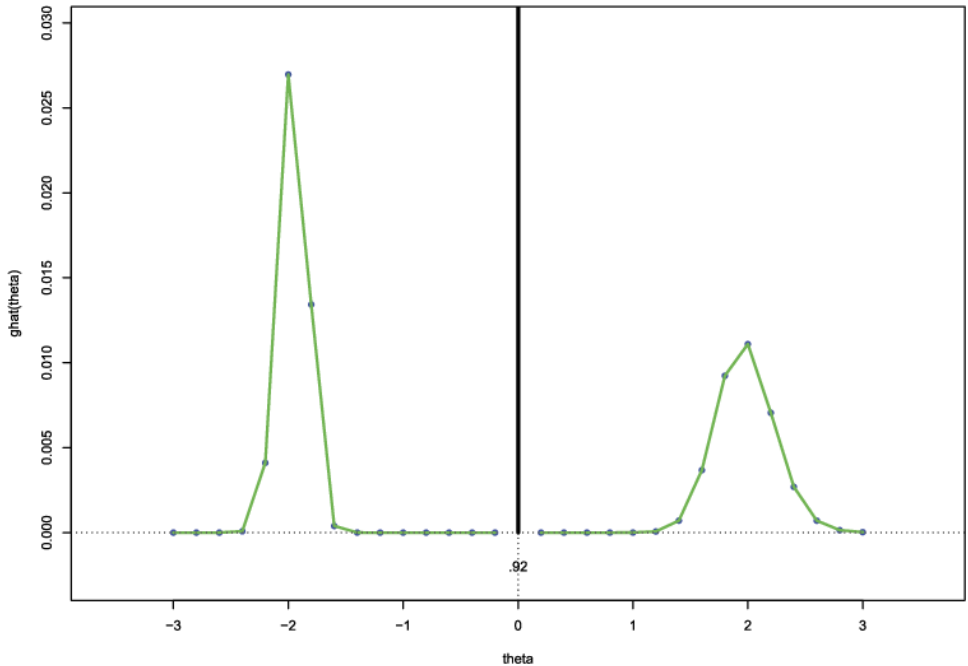}

\caption{MLE nonnull distribution, estimated from a sample of
$N=5000\ X$ values from $f$ corresponding to true $g$ in Figure~\protect\ref
{fig4}; estimated
atom at $\theta=0$ was 0.92.}
\label{fig6}
\end{figure*}

A random sample of $N=5000\ X$ values was drawn from the distribution
$\bmf=P\bg$ corresponding to the true $\bg$ in Figure~\ref{fig4} [with $P$
based on the normal density $\varphi(x_i-\theta_j)$ as before], giving
count vector $\by$ \eqref{37}. Numerical maximization yielded $\halp$,
the MLE in model \eqref{51}--\eqref{52}, $Q$ as in Table~\ref{tab3}. The
estimate $\hbg=\bg(\halp)$ put probability 0.920 at $\theta=0$,
compared to true value 0.903, with nonnull distribution as shown in
Figure~\ref{fig6}. The nonnull peaks at $\theta=\pm2$ were artifacts of the
estimation procedure. On the other hand, $\hbg$ correctly put roughly
equal nonnull probability above and below 0. This degree of useful but
crude inference\vadjust{\goodbreak} should be kept in mind for the genuine data examples of
Section~\ref{sec6}, where the truth is unknown.

Our list of $g$-modeling advantages raises the question of why
$f$-modeling has dominated empirical Bayes applications. The
answer---that a certain class of important problems is more naturally considered
in the $f$ domain---is discussed in the next section. Theoretically,
as opposed to practically, $g$-modeling has played a central role in
the empirical Bayes literature. Much of that work involves the
nonparametric maximum likelihood estimation of the prior distribution
$g\pthe$, some notable references being \citet{laird}, \citet{zhang} and
\citet{jiang}. Parametric $g$-modeling, as discussed in \citet{morris}
and \citet{casella}, has been less well developed. A large part of the
effort has focused on the ``normal-normal''\vadjust{\goodbreak} situation, normal priors
with normal sampling errors, as in \citet{1975morris}, and other
conjugate situations. Chapter~3 of \citet{newcarlin} gives a nice
discussion of parametric empirical Bayes methods, including binomial
and Poisson examples.

%s6 #&#
\section{Classic Empirical Bayes Applications}\label{sec6}

Since its post-war emergence (\cite{robbins}, \cite{good}, \cite{james}), empirical
Bayes methodology has focused on a small set of specially structured
situations: ones where certain Bayesian inferences can be computed
simply and directly from the marginal distribution of the observations
on the $x$-space. There is no need for $g$-modeling in this framework
or, for that matter, any calculation of $\hbg$ at all. False discovery
rates and the James--Stein estimator fall into this category, along
with related methods discussed in what follows. Though $g$-modeling is
unnecessary here, it will still be interesting to see how it performs
on the classic problems.

Robbins' Poisson estimation example exemplifies the classic empirical
Bayes approach: independent but not identically distributed Poisson variates
%
%e6.1 #&#
\begin{equation}
X_k\ind\operatorname{Poi}(\Theta_k),\quad k=1,2,\ldots,N,
\label{61}
\end{equation}
are observed, with the $\Theta_k$'s notionally drawn from some prior
$g\pthe$. Applying Bayes rule with the Poisson kernel $e^{-\theta
}\theta
^x/x!$ shows that
%
%e6.2 #&#
\begin{equation}
E\{\theta|x\}=(x+1)f_{x+1}/f_x, \label{62}
\end{equation}
where $\bmf=(f_1,f_2,\ldots)$ is the marginal distribution of the $X$'s.
[This is an example of \eqref{35}, Bayes rule in terms of $\bmf$;
defining $\be_i=(0,0,\ldots,1,0,\ldots,0)'$ with 1 in the $i$th place,
$\bU=(x+1)\be_{x+1}$, and $\bV=\be_x$.] Letting $\hbf=(\hatf
_1,\hatf
_2,\ldots)$ be the nonparametric MLE \eqref{310}, Robbins' estimate is
the ``plug-in'' choice
%
%e6.3 #&#
\begin{equation}
\hate\{\theta|x\}=(x+1)\hatf_{x+1}/\hatf_x \label{63}
\end{equation}
as in \eqref{311}. \citet{brown} use various forms of semiparametric
$f$-modeling to improve on \eqref{63}.

%f7 #&#
\begin{figure*}

\includegraphics{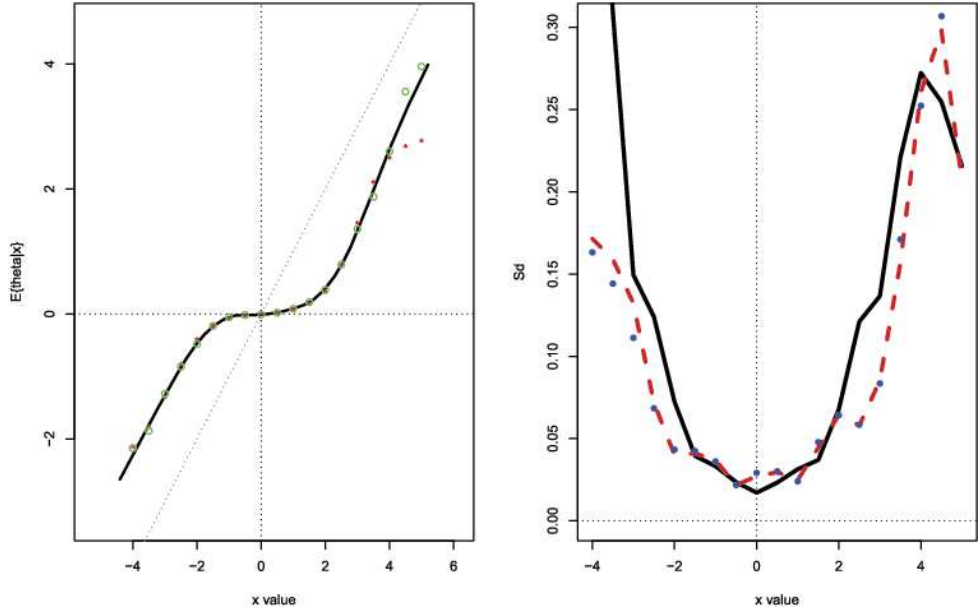}

\caption{\textit{Prostate data}. Left panel shows estimates of $E\{
\theta|x\}$ from Tweedie's formula (solid curve), $f$-modeling
(circles) and $g$-modeling (dots). Right panel compares standard
deviations of $\hate\{\theta|x\}$, for Tweedie estimates (dots),
$f$-modeling (dashed curve) and $g$-modeling (solid curve); reversals
at far right are computational artifacts.}
\label{fig7}
\end{figure*}

The prehistory of empirical Bayes applications notably includes the
\textit{missing species problem}; see Section~11.5 of \citet{2010}. This
has the Poisson form \eqref{61}, but with an inference different than
\eqref{62} as its goal. \citet{fisher} employed parameterized
$f$-modeling as in Section~\ref{sec4}, with $f$ the negative binomial family.
Section~3.2.1 of \citet{newcarlin} follows the same route for improving
Robbins' estimator \eqref{63}.\vadjust{\goodbreak}

\textit{Tweedie's formula} (\citep{2011}) extends Robbins-type estimation
of $E\{\theta|x\}$ to general exponential families. For the normal case
%
%e6.4 #&#
\begin{equation}
\theta\sim g\pdot\quad\mbox{and}\quad x|\theta\sim\caln(\theta,1),
\label{64}
\end{equation}
Tweedie's formula is
%
%e6.5 #&#
\begin{eqnarray}\label{65}
E\{\theta|x\}=x+l'(x)
\nonumber
\\[-8pt]
\\[-8pt]
\eqntext{\displaystyle\mbox{where }l'(x)=
\frac{d}{dx}\log f(x), }
\end{eqnarray}
with $f(x)$ the marginal distribution of $X$. As in \eqref{62}, the
marginal distribution of $X$ determines $E\{\theta|x\}$, without any
specific reference to the prior $g\pthe$.

Given observations $X_k$ from model \eqref{64},
%
%e6.6 #&#
\begin{equation}
X_k\sim\caln(\Theta_k,1)\quad\mbox{for }k=1,2,\ldots,N,
\label{66}
\end{equation}
the empirical Bayes estimation of $E\{\theta|x\}$ is conceptually
straightforward: a smooth estimate $\hatf(x)$ is obtained from the
$X_k$'s, and its logarithm $\hat{l}(x)$ differentiated to give
%
%e6.7 #&#
\begin{equation}
\hate\{\theta|x\}=x+\hat{l}'(x), \label{67}
\end{equation}
again without explicit reference to the unknown $g\pthe$. Modeling here
is naturally done on the $x$-scale. [It is not necessary for the
$X_k$'s to be independent in \eqref{66}, or \eqref{61}, although
dependence decreases the accuracy of $\hate$; see Theorem 8.4 of \citet{2010}.]

Figure~\ref{fig7} concerns an application of Tweedie's formula to the \textit
{prostate data}, the output of a microarray experiment comparing 52
prostate cancer patients with 50 healthy controls (\cite{2010}, Section~2.1). The genetic activity of $N=6033$ genes was
measured for each man. Two-sample tests comparing patients with
controls yielded $z$-values for each gene, $X_1,X_2,\ldots,X_N$,
theoretically satisfying
%
%e6.8 #&#
\begin{equation}
X_k\sim\caln(0,1) \label{68}
\end{equation}
under the null hypothesis that gene $k$ is equally active in both
groups. Of course, the experimenters were searching for activity
\textit
{differences}, which would manifest themselves as unusually large
values $|X_k|$. Figure~2.1 of \citet{2010} shows the histogram of the
$X_k$ values, looking somewhat like a long-tailed version of a $\caln
(0,1)$ density.

The ``smooth estimate'' $\hatf(x)$ needed for Tweedie's formula \eqref
{67} was calculated by Poisson regression, as in \eqref{43}--\eqref
{47}. The 6033 $X_k$ values were put into 193 equally spaced bins,
centered at $x_1,x_2,\ldots,x_{193}$, chosen as in \eqref{28} with $y_i$
being the number in bin $i$. A~Poisson generalized linear model \eqref
{43} then\vadjust{\goodbreak} gave MLE $\hbf=(\hatf_1,\hatf_2,\ldots,\hatf_{193})$.
Here the
structure matrix $\bX$ was the normal spline basis $ns(\mathbf{x},df=5)$
augmented with a column of 1's. Finally, the smooth curve $\hatf(x)$
was numerically differentiated to give $\hat{l}'(x)=\hatf'(x)/\hatf(x)$
and $\hate=x+\hat{l}'(x)$.

Tweedie's estimate $\hate\{\theta|x\}$ \eqref{67} appears as the solid
curve in the left panel of Figure~\ref{fig7}. It is nearly zero between $-2$
and 2, indicating that a large majority of genes obey the null
hypothesis \eqref{67} and should be estimated to have $\theta=0$. Gene
610 had the largest observed $z$-value, $X_{610}=5.29$, and
corresponding Tweedie estimate 4.09.

For comparison, $\hate\{\theta|x\}$ was recalculated both by
$f$-modeling as in Section~\ref{sec4} and $g$-modeling as in Section~\ref{sec5} [with
discrete sampling distributions \eqref{24}--\eqref{26} obtained from
$X_k\sim\caln(\Theta_k,1)$, $\Theta_k$ being the ``true effect size''
for gene $k$]; $f$-modeling used $\bX$ and $\hbf$ as just described,
giving $\hate_f=U_r'\hbf/V_r'\hbf$, $U_r$ and $V_r$ as in \eqref{419},
$r=12$; $g$-modeling took $\bthe=(-3,-2.8,\ldots,3)$ and $Q=(ns(\bthe
,5),\bone)$, yielding $\hbg=\bg(\halp)$ as the MLE from \eqref
{51}--\eqref{52}. [The R nonlinear maximizer \texttt{nlm} was used to
find $\halp$; some care was needed in choosing the control parameters
of \texttt{nlm}. We are paying for the fact that the $g$-modeling
likelihood \eqref{52} is not an exponential family.] Then the estimated
posterior expectation $\hate_g$ was calculated applying Bayes rule with
prior $\hbg$. Both $\hate_f$ and $\hate_g$ closely approximated the
Tweedie estimate.

Standard deviation estimates for $\hate_f$ [dashed curve, from \tref
{thm3} with $\hbf$ replacing $\bmf$ in \eqref{49}] and $\hate_g$ (solid
curve, from \tref{thm4}) appear in the right panel of Figure~\ref{fig7};
$f$-modeling gives noticeably lower standard deviations for $E\{\theta
|x\}$ when $|x|$ is large.

The large dots in the right panel of Figure~\ref{fig7} are bootstrap standard
deviations for the Tweedie estimates $\hate\{\theta|x\}$, obtained from
$B=200$ nonparametric bootstrap replications, resampling the $N=6033$
$X_k$ values. These closely follow the $f$-modeling standard
deviations. In fact, $\hate_f^*$, the bootstrap replications of $\hate
_f$, closely matched $\hate^*$ for the corresponding Tweedie estimates
on a case-by-case comparison of the 200 simulations. That is, $\hate_f$
is numerically just about the same as the Tweedie estimate, though it
is difficult to see analytically why this is the case, comparing
formulas \eqref{416} and \eqref{67}. Notice that the bootstrap results
for $\hate_f$ verify the accuracy of the delta-method calculations
going into \tref{thm3}.

%t4 #&#
\begin{table*}
\caption{Local false discovery rate estimates for the prostate data;
$\hufdr$ and its standard deviation estimates sdf obtained from
$f$-modeling; $\hfdr$ and sdg from $g$-modeling; sdf is substantially
smaller than sdg}\label{tab4}
\begin{tabular*}{\textwidth}{@{\extracolsep{\fill}}lccccccccc@{}}
\hline
$\bolds{x}$& $\bolds{-4}$& $\bolds{-3}$& $\bolds{-2}$& $\bolds{-1}$&
\textbf{0}& \textbf{1}& \textbf{2}& \textbf{3}& \textbf{4}\\
\hline
$\hufdr$&0.060& 0.370& 0.840&1.030&1.070&1.030& 0.860& 0.380& 0.050\\[3pt]
sdf& 0.014& 0.030& 0.034& 0.017& 0.013& 0.021& 0.033& 0.030& 0.009\\
sdg& 0.023& 0.065& 0.179& 0.208& 0.200& 0.206& 0.182& 0.068& 0.013\\[3pt]
$\hfdr$& 0.050& 0.320& 0.720& 0.880& 0.910& 0.870& 0.730& 0.320&
0.040\\
\hline
\end{tabular*}
\end{table*}

Among empirical Bayes techniques, the James--Stein estimator is
certainly best known. Its form,
%
%e6.9 #&#
\begin{eqnarray}\label{69}
\hthe=\bar{X}+ \bigl[1+(N-3)/S \bigr] (X_k-\bar{X} )
\nonumber
\\[-8pt]
\\[-8pt]
 \eqntext{\displaystyle\Biggl
[S=\sum
_1^N (X_k-\bar{X}
)^2 \Biggr],}
\end{eqnarray}
again has the ``classic'' property of being estimated directly from the
marginal distribution on the $x$-scale, without reference to $g\pthe$.
The simplest application of Tweedie's formula, taking $\bX$ in our
previous discussion to have rows $(1,x_i,x_i^2)$, leads to formula
\eqref{69}; see Section~3 of \citet{2011}.

Perhaps the second most familiar empirical Bayes applications relates
to \citeauthor{benjamini}'s (\citeyear{benjamini}) theory of false
discovery rates. Here we will
focus on the \textit{local false discovery rate} (fdr), which best
illustrates the Bayesian connection. We assume that the marginal
density of each observation of $X_k$ has the form
%
%e6.10 #&#
\begin{equation}
f(x)=\pi_0\varphi(x)+(1-\pi_0)f_1(x),
\label{610}
\end{equation}
where $\pi_0$ is the prior probability that $X_k$ is null, $\varphi(x)$
is the standard $\caln(0,1)$ density $\exp(-\frac{1}2 x^2)/\break \sqrt
{2\pi}$,
and $f_1(x)$ is an unspecified nonnull density, presumably yielding
values farther away from zero than does the null density $\varphi$.

Having observed $X_k$ equal to some value $x$, $\fdr(x)$ is the
probability that $X_k$ represents a null case \eqref{68},
%
%e6.11 #&#
\begin{equation}
\fdr(x)=\Pr\{\mathrm{null}|x\}=\pi_0\varphi(x)/f(x),
\label{611}
\end{equation}
the last equality being a statement of Bayes rule. Typically $\pi_0$,
the prior null probability, is assumed to be near 1, reflecting the
usual goal of large-scale testing: to reduce a vast collection of
possible cases to a much smaller set of particularly interesting ones.
In this case, the \textit{upper false discovery rate},
%
%e6.12 #&#
\begin{equation}
\ufdr(x)=\varphi(x)/f(x), \label{612}
\end{equation}
setting $\pi_0=1$ in \eqref{611}, is a satisfactory substitute for
$\fdr
(x)$, requiring only the estimation of the marginal density $f(x)$.

Returning to the discrete setting \eqref{29}, suppose we take the
parameter of interest $t\pthe$ to be
%
%e6.13 #&#
\begin{equation}
\bt=(0,0,\ldots,0,1,0,\ldots,0)', \label{613}
\end{equation}
with ``1'' at the index $j_0$ having $\theta_{j_0}=0$ [$j_0=16$ in~\eqref{27}].
Then $E\{t\pthe|x_i\}$ equals $\fdr(x_i)$, and we can
assess the accuracy of a $g$-model estimate $\hfdr(x_i)$ using \eqref
{518}, the corollary to \tref{thm4}.

This was done for the prostate data, with the data binned as in Figure~\ref
{fig7}, and $Q=(ns(\bthe,5),\bone)$ as before. \tref{thm4} was applied
with $\bthe$ as in \eqref{27}. The bottom two lines of Table~\ref{tab4} show
the results. Even with $N=6033$ cases, the standard deviations of
$\hfdr
(x)$ are considerable, having coefficients of variation in the 25\% range.

$F$-model estimates of fdr fail here, the bias/variance trade-offs of
Table~\ref{tab2} being unfavorable for any choice of $r$. However,
$f$-modeling is a natural choice for ufdr, where the only task is
estimating the marginal density $f(x)$. Doing so using Poisson
regression \eqref{43}, with $\bX=(ns(\mathbf{x},5),\bone)$, gave the
top two
lines of Table~\ref{tab4}. Now the standard deviations are substantially
reduced across the entire $x$-scale. [The standard deviation of $\hufdr
$ can be obtained from \tref{thm3}, with $\bU=\varphi(x_i)\bone$ and
$\bV$ the coordinate vector having 1 in the $i$th place.]

The top line of Table~\ref{tab4} shows $\hufdr(x)$ exceeding~1 near $x=0$.
This is the penalty for taking $\pi_0=1$ in~\eqref{612}. Various
methods have been used to correct $\hufdr$, the simplest being to
divide all of its values by their maximum. This amounts to taking $\hpi
_0=1/$maximum,
%
%e6.14 #&#
\begin{equation}
\hpi_0=1/1.070=0.935 \label{614}
\end{equation}
in Table~\ref{tab4}. [The more elaborate $f$-modeling program \texttt
{locfdr}, described in Chapter~6 of \citet{2010}, gave $\hpi_0=0.932$.]
By comparison, the $g$-model MLE $\hbg$ put probability $\hpi_0=0.852$
on $\theta=0$.

%t5 #&#
\begin{table}
\caption{$f$-modeling permits familiar and straightforward fitting
methods on the $x$ scale but then requires more complicated
computations for the posterior distribution of $\theta$; the situation
is reversed for $g$-modeling}\label{tab5}
\begin{tabular*}{\columnwidth}{@{\extracolsep{\fill}}lcc@{}}
\hline
&\multicolumn{1}{c}{\textbf{Model fitting}}&
\multicolumn{1}{c@{}}{\textbf{Bayesian computations}}\\
\hline
$f$-modeling&{direct}&{indirect}\\
$g$-modeling&{indirect}&{direct}\\
\hline
\end{tabular*}
\end{table}

%f8 #&#
\begin{figure*}[b]

\includegraphics{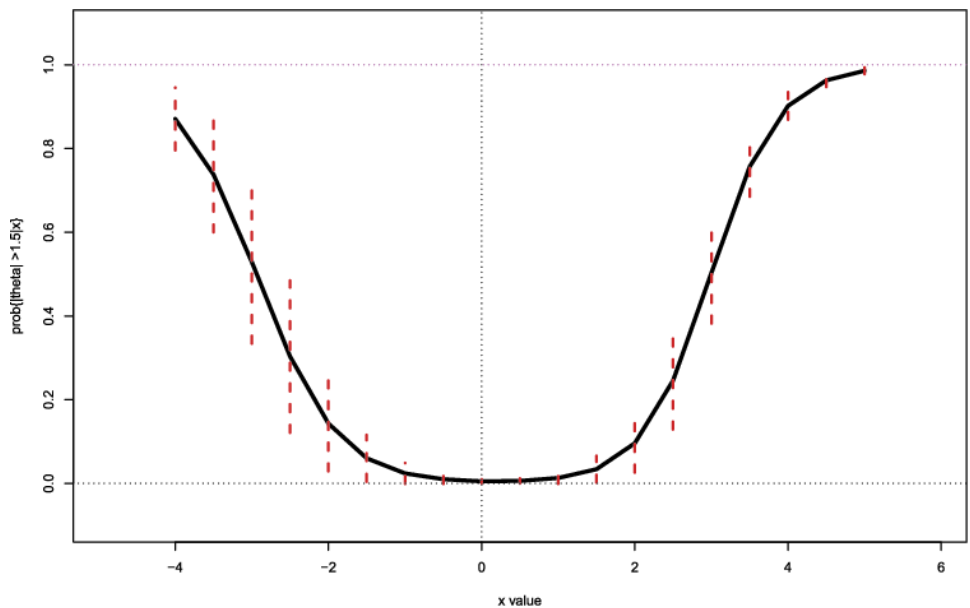}

\caption{$g$-modeling estimates of $\Pr\{|\theta|\geq1.5|x\}$ for the
prostate data. Dashed bars indicate $\pm$ one standard deviation, from
\protect\tref{thm4}.}
\label{fig8}
\end{figure*}

%s7 #&#
\section{Discussion}\label{sec7}

The observed data $X_1,X_2,\ldots,X_N$ from the empirical Bayes
structure \eqref{11}--\eqref{12} arrives on the $x$ scale but the
desired Bayesian posterior distribution $g(\theta|x)$ requires
computations on the $\theta$ scale. This suggests the two contrasting
modeling strategies diagrammed in Table~\ref{tab5}: modeling on the $x$
scale, ``$f$-modeling,'' permits the application of direct fitting
methods, usually various forms of regression, to the $X$ values, but
then pays the price of more intricate and less stable Bayesian
computations. We pay the price up front with ``$g$-modeling,'' where
models such as \eqref{52} require difficult nonconvex maximum
likelihood computations, while the subsequent Bayesian computations
become straightforward.

The comparative simplicity of model fitting on the $x$ scale begins
with the nonparametric case: $f$-modeling needs only the usual vector
of proportions $\hbf$ \eqref{310}, while $g$-modeling requires
\citeauthor{laird}'s (\citeyear{laird}) difficult nonparametric MLE
calculations. In general,
$g$-models have a ``hidden'' quality that puts more strain on
parametric assumptions; $f$-modeling has the advantage of fitting
directly to the observed data.

There is a small circle of empirical Bayes situations in which the
desired posterior inferences can be expressed as simple functions of
$f(x)$, the marginal distribution of the $X$ observations. These are
the ``classic'' situations described in Section~\ref{sec6}, and account for the
great bulk of empirical Bayes applications. The Bayesian computational
difficulties of $f$-modeling disappear here. Not surprisingly,
$f$-modeling dominates practice within this special circle.

``Bayes rule in terms of $f$,'' Section~\ref{sec2}, allows us to investigate
how well $f$-modeling performs outside the circle. Often not very well
seems to be the answer, as seen in the bottom panel of Figure~\ref{fig5}, for
example. $G$-modeling comes into its own for more general empirical
Bayes inference questions, where the advantages listed in Section~\ref{sec5}
count more heavily. Suppose, for instance, we are interested in
estimating $\Pr\{|\theta|\geq1.5|x\}$ for the prostate data. Figure~\ref{fig8}
shows the $g$-model estimates and their standard deviations from \tref
{thm4}, with $Q=ns(\bthe,6)$ as before. Accuracy is only moderate here,
but, nonetheless, some useful information has been extracted from the
data (while, as usual for problems involving discontinuities on the
$\theta$ scale, $f$-modeling is ineffective).

Improved $f$-modeling strategies may be feasible, perhaps making better
use of the kinds of information in Table~\ref{tab2}. A reader has pointed out
that pseudo-inverses of $P$ other than $A$ \eqref{31} are available, of
the form
%
%e7.1 #&#
\begin{equation}
\bigl(P'BP\bigr)^{-1}P'B. \label{71}
\end{equation}
Here the matrix $B$ might be a guess for the inverse covariance matrix
of $\hbf$, as motivated by generalized least squares estimation. So
far, however, situations like that in Figure~\ref{fig8} seem inappropriate for
$f$-modeling, leaving $g$-modeling as the only game in town.

Theorems \ref{th3} and \ref{th4} provide accuracy assessments for $f$-modeling and
$g$-modeling estimates. These can be dishearteningly broad. In the
bottom panel of Figure~\ref{fig5}, the ``good'' choice, $g$-modeling, would
still require more than $N=20\mbox{,}000$ independent observations $X_k$ to
get the coefficient of variation down to $0.1$ when $x$ exceeds 2. More
aggressive $g$-modeling, reducing the degrees of freedom for $Q$,
improves accuracy, at the risk of increased bias. The theorems act as a
reminder that, outside of the small circle of its traditional
applications, empirical Bayes estimation has an ill-posed aspect that
may call for draconian model choices. [The ultimate choice is to take
$g\pthe$ as known, that is, to be Bayesian rather than empirical
Bayesian. In our framework, this amounts to tacitly assuming an
enormous amount ``$N$'' of relevant past experience.]

Practical applications of empirical Bayes methodology have almost
always taken $\Theta_k$ and $X_k$ in \eqref{11}--\eqref{12} to be
real-valued, as in all of our examples. This is not a necessity of the
theory (nor of its discrete implementation in Section~\ref{sec2}). Modeling
difficulties mount up in higher dimensions, and even studies as large
as the prostate investigation may not carry enough information for
accurate empirical Bayes estimation.

There are not many big surprises in the statistics literature, but
empirical Bayes theory, emerging in the 1950s, had one of them: that
parallel experimental structures like \eqref{11}--\eqref{12} carry
within themselves their own Bayesian priors. Essentially, the other
$N-1$ cases furnish the correct ``prior'' information for analyzing
each $(\Theta_k,X_k)$ pair. How the statistician extracts that
information in an efficient way, an ongoing area of study, has been the
subject of this paper.

\section*{Acknowledgments}
I am grateful to Omkar Muralidharan, Amir
Najmi and Stefan Wager for many helpful discussions.
Research supported in part by NIH Grant 8R37 EB002784 and
by NSF Grant DMS-12-08787.

%
% imsref loaded by akundreckaite, 2013-12-20 13:00:46
%
% imsref loaded by akundreckaite, 2013-12-27 13:52:16

%

%suskaldyti doi

\end{document}